\documentclass[useAMS,usenatbib]{mn2e}

\usepackage{graphicx}

\title[Positron annihilation as a cosmic-ray probe]{Positron annihilation as a cosmic-ray probe}
\author[Y. Ohira, K. Kohri and N. Kawanaka]{Yutaka Ohira$^{1}$\thanks{E-mail:
ohira@post.kek.jp}, Kazunori Kohri$^{1}$ and Norita Kawanaka$^{2}$\\
$^{1}$Theory Center, Institute of Particle and Nuclear Studies, KEK, 1-1 Oho, Tsukuba 305-0801, Japan\\
$^{2}$Racah Institute of Physics, The Hebrew University, Jerusalem 91904, Israel}
\begin{document}

\date{Accepted 2011 December 28. Received 2011 December 27; in original form 2011 November 21}

\pagerange{\pageref{firstpage}--\pageref{lastpage}} \pubyear{2012}

\maketitle

\label{firstpage}

\begin{abstract}
Recently, the gamma-ray telescopes {\it AGILE\/} and {\it Fermi\/} observed several
middle-aged supernova remnants (SNRs) interacting with molecular clouds. 
A plausible emission mechanism of the gamma rays is the decay of neutral pions 
produced by cosmic ray (CR) nuclei (hadronic processes). 
However,  observations do not rule out contributions from bremsstrahlung 
emission due to CR electrons.
TeV gamma-ray telescopes also observed many SNRs and discovered many unidentified sources.
It is still unclear whether the TeV gamma-ray emission is produced via leptonic 
processes or hadronic processes.
In this Letter, we propose that annihilation emission of secondary positrons 
produced by CR nuclei is a diagnostic tool of the hadronic processes.
We investigate MeV emissions from secondary positrons and electrons produced 
by CR protons in molecular clouds.
The annihilation emission of the secondary positrons from SNRs can be robustly 
estimated from the observed gamma-ray flux. 
The expected flux of the annihilation line from SNRs observed by {\it AGILE\/} and 
{\it Fermi\/} is  sufficient for the future Advanced Compton Telescope to detect.
Moreover, synchrotron emission from secondary positrons and electrons and 
bremsstrahlung emission from CR protons can be also observed by the future 
X-ray telescope NuSTAR and ASTRO-H. 
\end{abstract}


\section{Introduction}
The origin of cosmic rays (CRs)  is a longstanding problem in astrophysics. 
Supernova remnants (SNRs) are thought to be the origin of Galactic CRs.
The most popular  SNR acceleration mechanism is diffusive shock 
acceleration \citep[e.g.][]{blandford87}.
In fact, {\it Fermi\/} and {\it AGILE\/} show that middle-aged SNRs 
($\sim10^4~{\rm yrs~old}$) interacting with molecular clouds emit 
GeV gamma rays with broken power law spectra
\citep[e.g.][]{abdo09,abdo10,tavani10,giuliani11} 
(Hereafter, these references are referred to as OBS). 
Considering a stellar wind before the supernova explosion, 
the molecular cloud has been swept away by the stellar wind.
The inner radius of the molecular cloud becomes typically about a few tens of 
parsecs. 
Then, the SNR collides with the molecular cloud at typically $10^4~{\rm yrs}$ 
later and the broken power law spectra of GeV gamma rays can be interpreted 
as the inelastic collision between molecular clouds and CR nuclei that have 
escaped from the SNR (hadronic processes) \citep*{ohiraetal11a}.
However, observations do not rule out contributions from bremsstrahlung 
emission due to CR electrons (leptonic processes). 
SNRs have been also observed by TeV gamma-ray telescopes, and 
it is still unclear whether the TeV gamma-ray emission is produced via 
inverse Compton scattering of CR electrons (leptonic processes) or 
hadronic processes.
There are many TeV-unID sources in our galaxy, and  their emission 
mechanisms are also still unclear.
Old or middle-aged SNRs are the candidates of the TeV-unID 
sources \citep{yamazaki06,ohiraetal11b}.
Thus, it is crucial to identify the emission mechanism of gamma rays.
Positrons and neutrinos are produced in hadronic processes.
On the other hand, they are not produced in leptonic processes.\footnote{
Strictly speaking, leptonic processes can produce positrons via 
$e^{-}+\gamma\rightarrow e^-+e^-+e^+$ but this reaction rate  
is suppressed by the fine structure constant compared with the gamma-ray production rate of inverse Compton scattering.}
Therefore, direct or indirect  detections of these particles will enable us 
to identify the emission mechanism in gamma-ray sources.

Recently, the CR positron excess has been provided 
by PAMELA \citep{adriani08}.
Although the origin of this excess has been investigated in many theoretical 
studies \citep[and references therein]{kashiyama11,kawanaka10,ioka10}, 
it still remains unclear.
SNRs have been discussed as the candidates of the positron sources 
\citep*{fujita09a,blasi09,shaviv09,biermann09,mertsch09}.
Hence, it is important to observe the positron production at SNRs.

Gamma-ray telescopes showed that the photon flux above $100\,{\rm MeV}$ 
from SNRs with an age of about 
$10^{4}\,{\rm yr}$ is $F_{>100\,{\rm MeV}}=\int_{100\,{\rm MeV}}^{\infty} {\rm d}E \left({\rm d}F/{\rm d}E\right)=10^{-7}-10^{-6}\,{\rm photon\,s^{-1}\,cm^{-2}}$ (OBS).
If the gamma rays have a hadronic origin, about the same numbers of 
positrons are produced.
When those positrons lose their energy sufficiently, about $80$ percent of 
them would make positroniums with ambient electrons, and one fourth of 
them would decay into two photons with the energy of $511\,{\rm keV}$ 
\citep[e.g.][]{prantzos10}.
This means that we can expect the $511\,{\rm keV}$ photon flux 
of the order of $10^{-8}-10^{-7}\,{\rm photon\,s^{-1}\,cm^{-2}}$.
This flux is sufficiently high for five-years observations of the future 
Advanced Compton Telescope (ACT) to detect.

In this Letter, considering the interaction between CR protons and 
molecular clouds, we estimate annihilation emission of secondary positrons 
produced by the CR protons.
We partly refer to analyses of previous works concerning the $511\,{\rm keV}$ 
line from the Galactic center \citep*{agaronyan81,beacom06,sizun06} and 
the Galactic plane \citep{stecker67,stecker69}, and we apply their treatments 
to middle-aged SNRs interacting with molecular clouds.
Recent review of positron annihilation can be found in \citet{prantzos10}.

We first provide some timescales for CR protons, secondary positrons 
and electrons inside a molecular cloud (Section \ref{sec:2}).
We then solve energy spectra of the secondary positrons and electrons 
(Section \ref{sec:3}) and calculate the annihilation line flux 
as well as the continuum spectrum  (Section \ref{sec:4}).
Section \ref{sec:5} is devoted to the discussion.

\section[]{Relevant timescales}
\label{sec:2}
\begin{figure}
\begin{center}
\includegraphics[width=80mm]{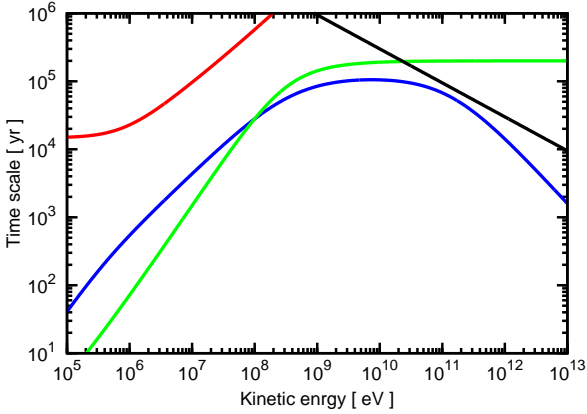}
\end{center}
\caption{Timescales of interaction for CRs in a molecular cloud with the number density $n=300\,{\rm cm^{-3}}$, the size $R_{c} = 30\,{\rm pc}$ and the magnetic field $B=30\,{\rm \mu G}$. 
The blue and green lines show the cooling times of secondary positrons (and electrons) and CR protons, respectively.
The red line shows the annihilation time of positrons.
The black solid line shows the diffusion time with $\chi = 0.01$.}
\label{fig1}
\end{figure}

In this section, we briefly summarize relevant timescales of CRs in a
molecular cloud. 
Physical quantities of molecular clouds associated with gamma-ray sources 
have not been understood in detail. 
Observations suggest that the gas number density is about 
$n=10^2-10^3\,{\rm cm^{-3}}$, the size is of the order of $10\,{\rm pc}$ (OBS), 
and the magnetic field in  molecular clouds is about $B=1-100\,{\rm \mu G}$ \citep{crutcher10}.
In this Letter, for the molecular cloud we adopt values of the gas number 
density $n=300\,{\rm cm^{-3}}$, the size $R_{c} = 30\,{\rm pc}$ and the 
magnetic field $B=30\,{\rm \mu G}$. 
We consider ionization and inelastic collisions with nuclei as cooling 
processes of CR protons.
On the other hand, we include ionization, bremsstrahlung, synchrotron 
emission, and inverse Compton scattering with the cosmic microwave 
background (CMB) as cooling processes of secondary positrons and 
electrons.
For SNRs observed by {\it Fermi\/} and {\it AGILE\/}, 
SNRs interact with molecular clouds 
and the expansion of the SNRs is strongly decelerated.
Therefore, we here do not consider the adiabatic cooling.

We here define $\tau_{\rm cool}= E/\dot{E}$ as a cooling time, where $E$ and
$\dot{E}$ are the kinetic energy and the energy loss rate, respectively. 
We adopt an expression of $\dot{E}$ for protons given in \citet{mannheim94}, 
and that for positrons and electrons given in \citet{strong98}. 
The timescale of direct annihilation of positrons and electrons is represented by
\begin{equation}
\tau_{\rm a,\pm}(E)=\frac{1}{n_{\rm \mp}\sigma_{\rm a}(E)v(E)}~~,
\end{equation}
where $n_{\rm \pm}, \sigma_{\rm a}$ and $v$ are the number density of positrons 
or electrons, the cross section of the annihilation \citep{dirac30} and the velocity of 
secondary electrons or positrons, respectively.
The electron density, $n_{-}$, is dominated by thermal electrons and the positron 
density, $n_{+}$, is very small, so that we can neglect the annihilation of secondary electrons. 

The escape time through  diffusion is written by
\begin{equation}
\tau_{\rm d}(E)=\frac{R_{\rm c}^2}{4D(E)}~~,
\end{equation}
where $D(E)$ is the diffusion coefficient. 
It is notable that so far the diffusion coefficient around an SNR has not been 
understood well. 
Thus we assume the diffusion coefficient in molecular clouds to be
\begin{equation}
D(E)=10^{28}~\chi \frac{v(E)}{c}\left( \frac{E}{10\,{\rm GeV}}\right)^{0.5} {\rm cm^2\,s^{-1}}~~,
\end{equation}
where $c$ is the velocity of light. 
Here $\chi=1$ might be reasonable as the Galactic mean value \citep{berezinskii90}, 
but $\chi=0.01$ should be preferred  in surroundings of SNRs 
\citep{fujita09b,fujita10,torres10,li10}, so that we use $\chi=0.01$ in this Letter.

In Fig.~\ref{fig1} we show those relevant timescales. 
The blue line shows the cooling time of secondary electrons and positrons 
where the synchrotron cooing dominates above $100\,{\rm GeV}$, 
the bremsstrahlung cooling dominates from $1\,{\rm GeV}$ to $100\,{\rm GeV}$, 
and the ionization loss dominates below $1\,{\rm GeV}$.
The green line shows the cooling time of CR protons where the pion production 
cooing dominate above $1\,{\rm GeV}$ and the ionization loss dominates below 1 GeV.
CR protons (about $1\,{\rm GeV}$) producing most positrons do not cool 
and escape as long as the SNR age is smaller than the cooling time 
of $1\,{\rm GeV}$ protons.
Cooling times of ionization, bremsstrahlung emission and inelastic collision 
are inversely proportional to the gas density. 
Therefore, spectral evolutions of all CRs below $1\,{\rm GeV}$ depends 
on only the density.
The black line shows the escape time due to diffusion with $\chi=0.01$.
The escape of secondary electrons and positrons is negligible compared 
with the energy loss. 

The typical energy of positrons produced by the CR protons is about 
$E_{\rm t}\sim 100\,{\rm MeV}$ \citep{murphy87}. 
Then its cooling time is about 
$3\times10^4\,(n/300\,{\rm cm}^{-3})^{-1}\,{\rm yr}$. 
Therefore, SNRs with an age longer than its cooling time can produce 
low energy positrons. 
Once positrons  have cooled to $100\,{\rm eV}$, they form positroniums 
through charge exchange. 
There are four possible spin configurations of the  positroniums. 
One of them has the total spin $0$ (singlet) and the three others have 
the total spin $1$ (triplet). 
The singlet state produces $511\,{\rm keV}$ line photons, and the triplet 
state produces continuum photons below $511\,{\rm keV}$. 
Because both the singlet and the triplet state are produced at the same rate, 
one fourth of positroniums can decay into two $511\,{\rm keV}$ line photons.

It is remarkable that the annihilation time is approximately $10$ times longer 
than the cooling time of positrons at around $10-100\,{\rm MeV}$. 
Then while the positron is being cooled, approximately $10$--$20$ percent of 
positrons directly annihilate with electrons. 
They produce continuum photons above $511\,{\rm keV}$.

\section[]{Energy spectra of secondary\\* positrons and electrons}
\label{sec:3}
\begin{figure}
\begin{center}
\includegraphics[width=80mm]{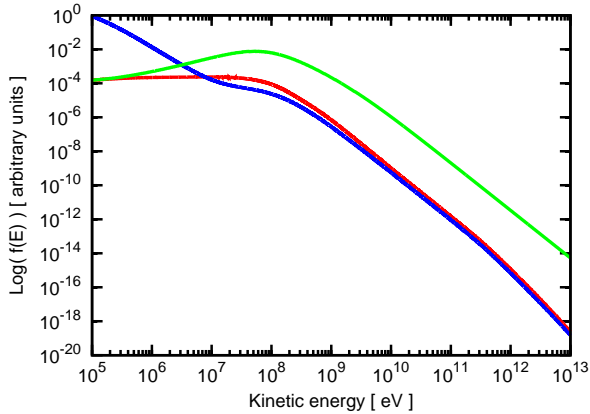}
\end{center}
\caption{Energy spectra of CR protons, secondary positrons and electrons for $s=2.8$ and $nt=2.8\times10^{14}\,{\rm cm^{-3}\,s}$. The green, red and blue lines show CR protons, secondary positrons and electrons spectra, respectively.}
\label{fig2}
\end{figure}

In this section, to calculate the photon spectrum from secondary positrons and electrons produced by CR protons, we calculate energy spectra of positrons and electrons, $f_{\pm}(t,E)$. 
An evolution of CR spectra is described by
\begin{equation}
\frac{\partial f}{\partial t}+\frac{\partial}{\partial E}\left(\dot{E}(E)f\right)+\frac{f}{\tau_{\rm a}(E)}+\frac{f}{\tau_{\rm d}(E)}=Q(t,E)~~,
\label{eq:df}
\end{equation}
where $Q(t,E)$ is the source term of CRs, the third and the fourth terms of
the left hand side are sink terms due to annihilation and diffusion, respectively. 
As shown in Fig.~\ref{fig1}, the diffusion escape for secondary electrons 
and positrons is negligible, so that we neglect the diffusion escape.
Then the solution for secondary positrons and electrons, $f_{\pm}$, can be expressed by
\begin{eqnarray}
f_{\pm}(t,E)&=&\frac{1}{|\dot{E}(E)|}\int_E^{E_0(t,E)}{\rm d} \epsilon ~Q_{\pm}(t',\epsilon) \nonumber \\ 
&&~~~~~~~~~~~~\times \exp\left(-\int_E^{\epsilon} \frac{{\rm d}\epsilon'}{\tau_{\rm a, \pm}|\dot{E}|}\right) ~~,
\label{eq:f}
\end{eqnarray}
where $t'$ is time for cooling from $E_0$ to $\epsilon$, and defined by
\begin{equation}
t'=\int_{\epsilon}^{E_0} \frac{{\rm d}\epsilon'}{|\dot{E}(\epsilon')|}=t-\int_{E}^{\epsilon} \frac{{\rm d}\epsilon'}{|\dot{E}(\epsilon')|}~~,
\end{equation}
and $E_0(t,E)$ is the initial energy of positrons or electrons before they cool to $E$ and defined by
\begin{equation}
\int_{E}^{E_0} \frac{{\rm d}\epsilon}{|\dot{E}(\epsilon)|}=t~~.
\end{equation}
We can neglect the annihilation loss for secondary electrons because 
the timescale is much longer than the SNR age.

Using the code provided by \citet{kamae06,karlsson08}, we calculate 
secondary positrons and electrons source spectra, $Q_{\pm}$, and the 
gamma-ray spectrum produced by decaying unstable hadrons such as pions and kaons.
In addition, for the high-energy electron source $Q_{-}$, we consider knock-on 
electrons produced by Coulomb collisions with CR protons \citep*{abraham66}.
The knock-on electrons contribute somewhat to the continuum emission as 
bremsstrahlung emission.

To obtain the source term of secondary positrons and electrons $Q_{\pm}(t,E)$, 
we need to calculate the spectral evolution of CR protons.
CR protons are injected from SNRs to molecular clouds in an 
energy-dependent way \citep*{ohiraetal10}.
Moreover, the diffusion escape from molecular clouds should be considered 
for CR protons above $10\,{\rm GeV}$ (see Fig.~\ref{fig1}).
However, we do not need to calculate the precise spectrum of CR protons 
above $1\,{\rm GeV}$ because thanks to {\it AGILE\/} and {\it Fermi\/}, 
we have already known spectra of CR protons above $1\,{\rm GeV}$. 
That is, we do not need to calculate the spectral evolution due to 
the diffusion escape for CR protons.
On the other hand, we have to calculate the evolution of the spectrum 
of CR protons below $1\,{\rm GeV}$ because gamma-ray spectra do not tell us 
the spectrum of CR protons below $1\,{\rm GeV}$.
Furthermore, the injection of CR protons is thought to stop when SNR 
shock collides with molecular clouds \citep{ohiraetal11a}, that is, 
the injection of CR protons stopped of the order of $10^4\,{\rm yrs}$ ago. 
Therefore, we assume that CR protons are injected as a delta function in time 
and an effective steep-spectrum, $Q_{\rm CR}$, 
instead of neglecting the diffusion escape
\begin{equation}
Q_{\rm CR}(t,E)= q_{\rm CR}(E)\delta(t)~~,
\end{equation}
where $q_{\rm CR}(E)$ is
\begin{equation}
q_{\rm CR}(E)\propto \left(E+m_{\rm p}c^2\right)\left\{E\left(E+2m_{\rm p}c^2\right)\right\}^{-\frac{1+s}{2}}~~,
\end{equation}
where $m_{\rm p}$ is the proton mass. 
This expression gives an injection term of $p^{-s}$, 
where $p$ is the momentum of the CR proton.
Observations show that spectra of CR protons have broken power-law forms 
and the spectral index above the break is $s=2.7-2.9$ except for W44 (OBS).
In this Letter, we adopt the single power law with $s=2.8$ 
(This assumption does not affect the flux of the annihilation line significantly).
Then, the solution to equation (\ref{eq:df}) for CR protons, $f_{\rm CR}$, can be expressed by 
\begin{equation}
f_{\rm CR}(t,E)= \frac{\dot{E}(E_0(t,E))}{\dot{E}(E)}q_{\rm CR}(E_0(t,E))~~.
\label{eq:fcr}
\end{equation}
As mentioned in previous section, the cooling of CRs below $1\,{\rm GeV}$ 
depends on only the density.
Hence, the solution (equation (\ref{eq:fcr})) depends on only $nt$.

To calculate source terms of secondary electrons and positrons, $Q_{\pm}(t',E)$, 
in equation (\ref{eq:f}), 
we approximately use the present spectrum of CR protons, $f_{\rm CR}(t,E)$, 
by changing $Q_{\pm}(t',E)$ into $Q_{\pm}(t,E)$ in equation~(\ref{eq:f}).
This is because CR protons (about $1\,{\rm GeV}$) producing most positrons do 
not cool and escape for SNRs observed by {\it Fermi\/} and {\it AGILE\/}.
The solution for the secondary positrons and electrons 
(equation~(\ref{eq:f})) depends on only $nt$ below $100\,{\rm GeV}$.
Hereafter, we use $nt$ as the parameter to describe the system 
(for example, $nt=2.8\times10^{14}\,{\rm cm^{-3}\,s}$ for 
$n=300\,{\rm cm}^{-3}$ and $t=3\times 10^4\,{\rm yr}$).
After the cooling time of CR protons above $1\,{\rm GeV}$ 
($nt>1.9\times10^{15}\,{\rm cm^{-3}\,s}$), all CRs have already cooled 
and we do not expect any emission.

In Fig.~\ref{fig2} we show the energy spectra of the CR protons,
secondary positrons and electrons given in equations (\ref{eq:f}) and (\ref{eq:fcr}), where $s=2.8, nt=2.8\times10^{14}\,{\rm cm^{-3}\,s}$.
For CR protons (green line), the spectrum below $100\,{\rm MeV}$ is modified from the initial spectrum $q_{\rm CR}(E)$ by ionization. 
The electron spectrum (blue lines) is dominated by knock-on electrons below $10\,{\rm MeV}$.

A steady-state solution is obtained by changing $E_0$ to infinity in equation (\ref{eq:f}). 
The steady-state spectrum of positrons below the typical energy of the source term $E_{\rm t}$ is approximately obtained by using the following approximation.
\begin{equation}
Q_+(E)= q_+\delta(E-E_{\rm t})~~,
\end{equation}
where $q_+$ is the total production number of the secondary positrons per unit time.
Then, the steady-state spectrum below $E_{\rm t}$ is given by 
\begin{equation}
f_{+}(E)=\frac{q_{+}}{|\dot{E}(E)|}\exp\left(-\int_E^{E_{\rm t}} \frac{{\rm d}\epsilon'}{\tau_{\rm a, +}|\dot{E}|}\right)~~.
\label{eq:fss}
\end{equation}
%

\section{Radiation spectrum}
\label{sec:4}
\begin{figure}
\begin{center}
\includegraphics[width=80mm]{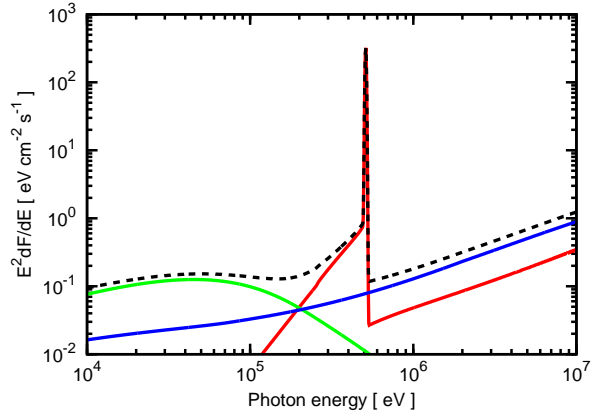}
\end{center}
\caption{Photon spectrum from an SNR with $F_{>100\,{\rm MeV}}=10^{-6}\,{\rm photon\,s^{-1}\,cm^{-2}}, nt=2.8\times 10^{14}\,{\rm cm^{-3}\,s}$ and $s=2.8$. 
The black dashed line shows the total spectrum. 
The red, blue and green lines show the annihilation spectrum, the bremsstrahlung spectrum from secondary positrons and electrons, and the bremsstrahlung spectrum from CR protons, respectively.}
\label{fig3}
\end{figure}
\begin{figure}
\begin{center}
\includegraphics[width=80mm]{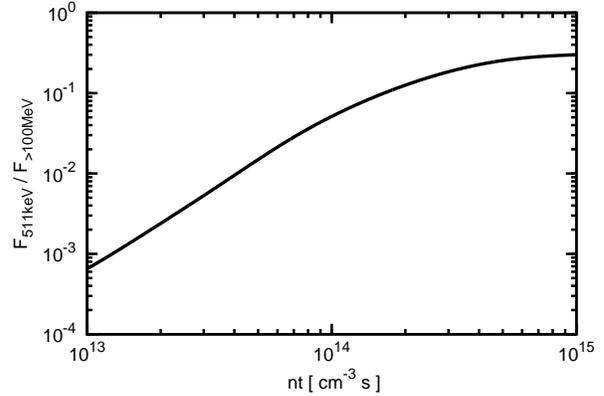}
\end{center}
\caption{The line shows $F_{511\,{\rm keV}}/F_{>100\,{\rm MeV}}$ as a function of $nt$ for $s=2.8$.}
\label{fig4}
\end{figure}

In this section, we calculate the radiation spectrum from secondary positrons 
and electrons at $nt=2.8\times10^{14}\,{\rm cm^{-3}\,s}$. 
In this case, primary CR electrons are negligible as long as the ratio of primary 
CR electrons to CR protons is $K_{\rm ep} < 0.01$ (see Fig.~\ref{fig2}).
We calculate synchrotron emission, inverse Compton emission with CMB 
and bremsstrahlung emission from secondary positrons 
and electrons \citep*{strong00}, bremsstrahlung emission from CR 
protons \citep{schuster03}, and annihilation emission 
from secondary positrons \citep[e.g.][]{sizun06}. 
Energy spectra of secondary positrons and electrons, obtained from 
equation (\ref{eq:f}), are shown in Fig.~\ref{fig2}.

When positrons cool to $\sim 10^2\,{\rm eV}$, they form positroniums 
through charge exchange. 
The production rate of the positroniums, $Q_{\rm Ps}$, is obtained by
\begin{equation}
Q_{\rm Ps}(t)=|\dot{E}(100~{\rm eV})|f_{+}(t,100~{\rm eV})~~.
\label{eq:qps}
\end{equation}
One fourth of positroniums decay into two $511\,{\rm keV}$ photons. 
Thus the photon flux of the $511\,{\rm keV}$ line, $F_{511\,{\rm keV}}$, is given by
\begin{equation}
F_{511{\rm keV}}(t)=\frac{Q_{\rm Ps}(t)}{8\pi d^2}~~.
\label{eq:f511}
\end{equation}
where $d$ is the source distance. 
The line width of the annihilation line of positroniums emitted from 
molecular clouds is $6.4\,{\rm keV}$ \citep*{guessoum05}. 
We here use a Gaussian profile with the $6.4\,{\rm keV}$ width 
as the line structure.

Fig.~\ref{fig3} shows the photon spectrum normalized so as to make 
$F_{>100\,{\rm MeV}}=10^{-6}\,{\rm photon\,s^{-1}cm^{-2}}$, 
where $nt=2.8\times 10^{14}\,{\rm cm^{-3}\,s}$ and $s=2.8$. 
We again note that this condition is typical for middle-aged SNRs observed 
by gamma-ray telescopes. 
Annihilation emission (red line) dominates at around the $511\,{\rm keV}$ range. 
It is notable that synchrotron emission and inverse Compton emission 
with CMB induced by secondary positrons and electrons do not contribute 
in the energy range of Fig.~\ref{fig3}. 
Moreover, synchrotron emission by secondary positrons and electrons 
depends on the magnetic field, the maximum energy of CR protons 
and the spectral index of CR protons ($s$). 
Bremsstrahlung emission of CR protons (green line) also depends on $s$.
Note that bremsstrahlung emission of CR protons can be observed 
by the future X-ray telescope, NuSTAR \citep{harrison05} and 
ASTRO-H \citep{takahashi10}.

For the estimation of the annihilation line,
we can neglect the escape loss as long as $\tau_{\rm cool}<\tau_{\rm d}$ 
for $E<100\,{\rm MeV}$.
Moreover, most positrons are cooled by the ionization loss as long as 
$B<1\,{\rm mG}\,(n/300\,{\rm cm^3})^{1/2}$, so that the magnetic field ($B$), 
the size of molecular clouds ($R_{\rm c}$) and 
the normalization of the diffusion coefficient ($\chi$) are not important.
Hence, the ambiguity of the annihilation line is only $nt$.  
Fig.~\ref{fig4} shows the flux ratio, $F_{511\,{\rm keV}}/F_{>100\,{\rm MeV}}$, 
as a function of $nt$, where $s=2.8$. 
The ratio does not depend on the number of CRs and the distance.
The flux ratio, $F_{511\,{\rm keV}}/F_{>100\,{\rm MeV}}$, 
becomes constant after 
$nt=2.8\times 10^{14}\,{\rm cm^{-3}\,s}$ because almost all positrons have 
cooled to $100~{\rm eV}$ by that time.
From Fig.~\ref{fig4} we expect $F_{511\,{\rm keV}} \sim 10^{-7}\,
{\rm photon\,cm^{-2}\,s^{-1}}$ from SNRs with $nt=10^{14}-10^{15}\,
{\rm cm^{-3}\,s}$ and with $F_{>100\,{\rm MeV}}\sim10^6\,
{\rm photon\,cm^{-2}\,s^{-1}}$. 
The expected flux of the annihilation line, $F_{511\,{\rm keV}}$,
can be sufficient for five-years observations of ACT to detect \citep{boggs06}.
Therefore we can expect to detect the annihilation line of secondary 
positrons produced by the CR protons in SNRs observed by {\it AGILE\/} 
and {\it Fermi\/}.

\section{Discussion and Summary}
\label{sec:5}

In this Letter, we  have investigated the MeV emission spectrum 
due to CR protons from SNRs interacting with molecular clouds.
We found that for typical middle-aged SNRs observed by {\it AGILE\/} 
and {\it Fermi\/}, secondary positrons can cool to an energy sufficient 
to make positroniums.
We calculated annihilation emission of the positrons and other emissions. 
The expected flux of the annihilation line is sufficient for the future 
gamma-ray telescope, such as ACT, to detect.
Therefore, we propose that annihilation emission from secondary positrons 
is a important tool as a CR probe.
Moreover, synchrotron emission from secondary positrons and electrons, 
and bremsstrahlung emission from CR protons can be also observed 
by the future X-ray telescope, NuSTAR \citep{harrison05} and 
ASTRO-H \citep{takahashi10}.

All particles with energies smaller than $1\,{\rm GeV}$ lose their energy due 
to ionization, as shown in Fig.~\ref{fig1}.
Not only the $511\,{\rm keV}$ line but also the ionization rate is also a probe 
of CR nuclei \citep{goto08,indriolo10}.
\citet{becker11} proposed that ${\rm H}_2^{+}$ and ${\rm H}_3^{+}$ lime emissions
should be observed from molecular clouds if there are many CRs. 
Low energy CR protons and knock-on electrons might be measured by 
using atomic lines \citep{gabriel79,tatischeff03}, so that atomic lines also 
become a probe of CR nuclei.
Quantitative estimations will be addressed in future work.
Moreover, CR nuclei can produce nuclear excitation lines by inelastic collisions 
and productions of unstable nuclei \citep*[e.g.][]{nath94,tatischeff03,summa11}.
Therefore, CR compositions around SNRs can be also investigated 
by the nuclear excitation lines.
Recent observations and theoretical studies for CR compositions are also 
remarkable \citep{ahn10,ohira11}

The spectra of secondary positrons and electrons produced by CR protons 
are different from a single power law because of their cooling and injection 
spectra (see Fig.~\ref{fig2}).
The spectra of secondary positrons and electrons below $100\,{\rm MeV}$ 
would become harder than the CR proton spectrum around $1\,{\rm GeV}$. 
The number of secondary positrons and electrons can be larger than that 
of primary CR electrons when the SNR age is larger than about 10 percent 
of the cooling time of CR protons due to the inelastic collision.
Even when the number of the primary CR electrons is larger than that of 
secondary  positrons and electrons, the spectrum is affected by the cooling.
These cooling effects of positrons and electrons might be a reason why radio  
spectra are different from expected from gamma-ray  spectra observed 
by {\it Fermi\/} \citep{uchiyama10}. 
However, the radio emission from SNRs may be originated from 
inside the SNR, but observed gamma rays may be originated from 
outside the SNR.
That is, a different component may produce the radio emission.
We may have to build more detailed models to compare the theory and 
observed date in future.
It is an interesting future work to compare between observations and 
the theory for individual SNRs. 

\section*{Acknowledgments}

We thank K. Ioka for useful comments.
This work is supported in part by grant-in-aid from the Ministry of Education, Culture, Sports, Science, and Technology (MEXT) of Japan, No.~21684014 (Y.~O.), No.~22740131 (N.~K.).
K.K. was partly supported by the Center for the Promotion of
Integrated Sciences (CPIS) of Sokendai, and Grant-in-Aid for
Scientific Research on Priority Areas
No. 18071001, Scientific Research (A) No.22244030 and Innovative Areas
No. 21111006.

\bsp
\label{lastpage}
\end{document}